%% file: paper.tex
\documentclass[a4paper,USenglish,cleveref]{lipics-v2021}

\hideLIPIcs
\pdfoutput=1
\usepackage{tikz}
\usetikzlibrary{positioning}

\title{Wavelet Forests Revisited}

\author{Eric Chiu}{Department of Computer Science, Stony Brook University, NY, USA}{echiu@cs.stonybrook.edu}{https://orcid.org/0009-0003-4962-3436}{}
\author{Dominik Kempa}{Department of Computer Science, Stony Brook University, NY, USA}{kempa@cs.stonybrook.edu}{https://orcid.org/0000-0003-2286-7417}{}

\authorrunning{E. Chiu and D. Kempa}
\Copyright{Eric Chiu and Dominik Kempa}


\supplementdetails{Source Code}{https://github.com/echiu12/wavelet-forest}
\funding{Partially funded by the NSF CAREER Award 2337891.}
\nolinenumbers

\EventEditors{Martin Aum\"{u}ller and Irene Finocchi}
\EventNoEds{2}
\EventLongTitle{24th Symposium on Experimental Algorithms (SEA 2026)}
\EventShortTitle{SEA 2026}
\EventAcronym{SEA}
\EventYear{2026}
\EventDate{June 22--24, 2026}
\EventLocation{Copenhagen, Denmark}
\EventLogo{}
\SeriesVolume{371}
\ArticleNo{22}
\input{settings}

\input{macros}

\begin{document}

\maketitle

\input{abstract}

\input{intro}

\input{prelim}

\input{wavelet-forest}

\input{implementation}

\input{experiments}

\bibliographystyle{plainurl}
\bibliography{paper}

\end{document}

%% file: settings.tex
\hypersetup{
  bookmarksopen,
  bookmarksnumbered,
}

\allowdisplaybreaks

%% file: macros.tex
\newcommand{\bigO}{\mathcal{O}}
\newcommand{\dd}{\mathinner{.\,.}}
\newcommand{\dol}{\text{\normalfont $\texttt{\$}$}}
\newcommand{\Z}{\mathbb{Z}}

\newcommand{\Rank}[3]{\operatorname{rank}_{#1}(#2,#3)}
\newcommand{\Select}[3]{\operatorname{select}_{#1}(#2,#3)}

%% file: abstract.tex
\begin{abstract}
  Rank and select queries are basic operations on sequences, with
  applications in compressed text indexes and other space-efficient
  data structures. One of the standard data structures supporting
  these queries is the \emph{wavelet tree}. In this paper, we study
  \emph{wavelet forests}, that is, wavelet-tree structures based on
  the \emph{fixed-block compression boosting} technique.
  Such structures partition the input
  sequence into fixed-size blocks and build a separate wavelet tree
  for each block. Previous work showed that this approach yields
  strong practical performance for rank queries.

  We extend wavelet forests to support select queries. We show that
  select support can be added with little additional space overhead
  and that the resulting structures remain practically efficient. In
  experiments on a range of non-repetitive and repetitive inputs,
  wavelet forests are competitive with, and in most cases outperform,
  standalone wavelet-tree implementations. We also study the effect
  of internal parameters, including superblock size and navigational
  data, on select-query performance.
\end{abstract}

%% file: intro.tex
\section{Introduction}\label{sec:intro}

Consider a string $S \in [0 \dd \sigma)^{n}$ of length $n$ over the
integer alphabet $[0 \dd \sigma)$.
\begin{itemize}
\item Given any $i \in [0 \dd n]$ and any symbol $c \in [0 \dd
  \sigma)$, the \emph{rank query} returns the value $\Rank{S}{i}{c} :=
  |\{j \in [1 \dd i] : S[j] = c\}|$, i.e., the number of occurrences
  of $c$ in $S[1 \dd i]$;
\item Given any $r \in \Z_{\geq 1}$ and any symbol $c \in [0 \dd
  \sigma)$, the \emph{select query} returns the value
  $\Select{S}{r}{c}$, defined as the $r$th smallest element of the set
  $\{i \in [1 \dd n] : S[i] = c\}$. If the size of this set is less
  than $r$, we set $\Select{S}{r}{c} = \infty$.
\end{itemize}

Rank and select queries are among the most basic queries on
sequences. They form the backbone of many compressed data structures
for strings~\cite{FerraginaM05,GrossiV05,%
  Sadakane07,rindex,%
  KempaK19,KempaK23,resolution}.
They are also core components of space-efficient representations of
more complex data types, including permutations, parentheses, document
collections, trees, graphs, and
grids~\cite{navarrobook,Navarro14,FerraginaGM09,Grossi16a,GagieNP12}.

One of the most efficient data structures supporting rank and select
queries is the \emph{wavelet tree}, invented by Grossi, Gupta, and
Vitter~\cite{wavelet_tree}. For a string $S \in [0 \dd \sigma)^{n}$,
wavelet trees use $\bigO(n \log \sigma)$ bits and support rank and
select queries in $\bigO(\log \sigma)$ time.\footnote{Space of wavelet trees
can also be bounded in terms of the
$k$th-order empirical entropy $H_k(S)$ of
$S$~\cite{wavelet_tree}.} Efficient implementations of wavelet trees
have been the subject of intense study, and multiple highly efficient
solutions have been designed, optimizing many aspects of wavelet
trees, including space usage, query time, construction time, and
working space, both in theory and in
practice~\cite{wavelet_matrix,Gog11,GrossiVX11,%
  Tischler11,BabenkoGKS15,MunroNV16,sdsl,%
  LabeitSB17,Shun20,DinklageEFKL21,%
  Dinklage0K20,DinklageFKT23,CereginiKV24,FerraginaGGRVV25,QWT}.

In this paper, we focus on a particular wavelet-tree implementation
called a \emph{wavelet forest}~\cite{wavelet_forest}.\footnote{The
  technique introduced in~\cite{wavelet_forest} is also called
  \emph{fixed-block compression boosting}.} The basic idea behind
wavelet forests is to partition the input sequence into blocks and
construct a wavelet tree for each block, while also maintaining global
rank values at block boundaries; these boundaries are then grouped
into larger hierarchies such as superblocks and hyperblocks. Simply
splitting the sequence and constructing a \emph{standalone} wavelet
tree for every block---even using state-of-the-art wavelet-tree
implementations---does not yield a significant improvement. However,
it is shown in~\cite{wavelet_forest} that, when this idea is combined
with several algorithmic improvements (including alphabet mapping at
the block and superblock levels, merging the bitvectors of all
wavelet trees into one, sharing lookup tables, dynamic selection of
blocks in each superblock, pointerless tree navigation, etc.), wavelet
forests achieve excellent performance, leading to new state-of-the-art
results for rank queries. A unique advantage of wavelet forests is
their \emph{adaptability}: they achieve strong practical performance
regardless of the underlying sequence type. In particular, they
work well on both \emph{non-repetitive} and \emph{highly repetitive}
input texts, making them an excellent off-the-shelf choice in many
applications. The implementation of the wavelet forest
from~\cite{wavelet_forest} is part of the \texttt{SDSL}
library---a powerful and flexible library of succinct data
structures~\cite{sdsl}.\footnote{Available at
  \url{https://github.com/simongog/sdsl-lite}.} However, the
implementation of the wavelet forest in~\cite{wavelet_forest} supports
only rank queries and lacks support for select queries. Until now, it
has not been known whether the strong practical performance of wavelet
forests also extends to select queries, and, if so, at what cost.

\subparagraph{Our Results}

The contribution of this paper is twofold:
\begin{itemize}
\item We propose the first efficient implementation of select queries
  on wavelet forests. With only minor additions, wavelet forests
  support select queries with little extra cost in practice and, in
  most cases, outperform standalone optimized wavelet trees. For
  example, with the RRR bitvector implementation, they use the same or
  less space while improving query time by up to a factor of two.
\item In addition to demonstrating the performance of our structure on
  a range of inputs (\cref{sec:experiments-exp-1}), we also
  explore the effect of internal parameters, including the superblock
  size and the impact of navigational headers
  (\cref{sec:experiments-exp-2,sec:experiments-exp-3}).
\end{itemize}

\subparagraph{Organization of the Paper}

In \cref{sec:prelim}, we introduce the notation and definitions used
throughout the paper. In \cref{sec:wavelet-forest}, we present a basic
overview of the components of wavelet forests. In
\cref{sec:implementation}, we describe the details of our
implementation. Finally, in \cref{sec:experiments}, we present
experiments demonstrating the performance of wavelet
forests for select queries.

%% file: prelim.tex
\section{Preliminaries}\label{sec:prelim}

\subparagraph{Strings}

A \emph{string} is a finite sequence of symbols from a given set
$\Sigma$, called the \emph{alphabet}. For any $i \in [1 \dd |S|]$, we
denote the $i$th leftmost symbol of $S$ by $S[i]$. Strings of the form
$S[i \dd j)$, where $1 \leq i \leq j \leq |S|+1$, are called
\emph{substrings} or \emph{factors} of $S$.\footnote{We use $[i \dd j)$,
  $(i \dd j)$, and $(i \dd j]$ as shorthand for $[i \dd j-1]$,
  $[i+1 \dd j-1]$, and $[i+1 \dd j]$, respectively.}
When $i = 1$ (resp.\ $j = |S| + 1$), the substring $S[i \dd j)$ is a
\emph{prefix} (resp.\ \emph{suffix}) of $S$. The concatenation of
strings $S_1$ and $S_2$ is denoted by $S_1S_2$ or $S_1 \cdot S_2$. We
assume that the set $\Sigma$ is equipped with an order denoted by
$\preceq$. The \emph{lexicographic order} on strings over $\Sigma$ is
the extension of the order $\preceq$ on $\Sigma$ defined as follows:
for strings $S_1$ and $S_2$, it holds that $S_1 \preceq S_2$ if either
$S_1$ is a prefix of $S_2$, or there exists $\ell \in [1 \dd
\min(|S_1|, |S_2|)]$ such that $S_1[1 \dd \ell) = S_2[1 \dd \ell)$ and
$S_1[\ell] \prec S_2[\ell]$.

\subparagraph{Burrows--Wheeler Transform}

Consider a string $S \in \Sigma^{n}$ such that $S[n]$ is a unique
symbol in $S$, denoted by $S[n] = \dol$, and $\dol$ is the smallest
symbol in $\Sigma$. Let $\mathcal{M}$ denote the $n \times n$ matrix
obtained by lexicographically sorting all rotations of $S$, that is,
all strings in the set $\{S[i \dd n] \cdot S[1 \dd i{-}1] : i \in [1
\dd n]\}$. The \emph{Burrows--Wheeler Transform (BWT)} of $S$ is the
string formed by taking the last column of $\mathcal{M}$~\cite{bwt}.

\begin{example}\label{ex:bwt}
  The BWT of the string $S = \texttt{BANANA\dol}$ is
  $\texttt{ANNB\dol AA}$.
\end{example}

\subparagraph{Wavelet Trees}

The wavelet tree $\mathcal{T}$ of $S \in \Sigma^{n}$ is a binary tree
with $\sigma = |\Sigma|$ leaves, each representing a symbol in the
alphabet~\cite{wavelet_tree}.
Every internal node $v$ of $\mathcal{T}$ represents a subset
$\Sigma_{v} \subseteq \Sigma$ of symbols corresponding to the leaves in
the subtree rooted at $v$. Each node $v$ has an associated
string $S_v$, which is a subsequence of $S$ containing only symbols in
$\Sigma_{v}$. Lastly, each internal node has an associated bitvector
$B_v$ containing, for every position $j \in [1 \dd |S_v|]$, information
indicating whether $S_v[j]$ is represented by a leaf in the subtree
rooted at the left or right child of $v$. Each bitvector $B_v$ is
augmented with a data structure supporting rank and select queries over
the binary alphabet. Assuming these queries take $\bigO(1)$ time,
wavelet trees support rank and select queries over $S$ in $\bigO(\log
\sigma)$ time~\cite{wavelet_tree}.

One of the central applications of wavelet trees is supporting rank
queries over the BWT of a given text. Such functionality is at the core
of the FM-index~\cite{FerraginaM05} and enables efficient pattern
matching over the underlying text. In \cref{fig:wt}, we show the wavelet
tree for the BWT of the text from Example~\ref{ex:bwt}.

\input{figs/wt}

%% file: figs/wt.tex
\begin{figure}
	\centering
	\begin{tikzpicture}[node distance=1mm]
		\node[draw,circle] (A) at (0, 0) {};
		\begin{scope}[node distance=8mm and 12mm]
			\node[draw,circle,below left=of A]  (B) {};
			\node[draw,circle,below right=of A] (C) {};
		\end{scope}
		\begin{scope}[node distance=8mm and 3mm]
			\node[draw,below left=of B]  (D) {\texttt{\$}};
			\node[draw,below right=of B] (E) {\texttt{A}};
			\node[draw,below left=of C]  (F) {\texttt{B}};
			\node[draw,below right=of C] (G) {\texttt{N}};
		\end{scope}
		\draw (A) -- (B);
		\draw (A) -- (C);
		\draw (B) -- (D);
		\draw (B) -- (E);
		\draw (C) -- (F);
		\draw (C) -- (G);
		\node[align=center,right=of A] {\texttt{ANNB\$AA}\\[-2mm]\texttt{0111000}};
		\node[align=center,left=of B] {\texttt{A\$AA}\\[-2mm]\texttt{1011}};
		\node[align=center,right=of C] {\texttt{NNB}\\[-2mm]\texttt{110}};
	\end{tikzpicture}
	\caption{An illustration of a wavelet tree for the string \texttt{ANNB\$AA}.}
	\label{fig:wt}
\end{figure}
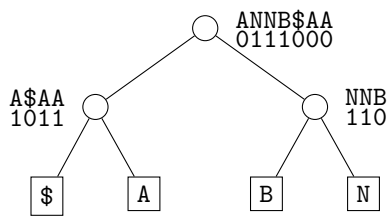

%% file: wavelet-forest.tex
\section{Wavelet Forest}\label{sec:wavelet-forest}

In this section, we describe the basic idea behind the wavelet
forest~\cite{wavelet_forest} and introduce the notation used in the
subsequent sections. At a high level, a wavelet forest represents the
input text by partitioning it into blocks and building a separate
wavelet tree for each block. This approach is justified theoretically
by the \emph{fixed-block boosting} theorem proved
in~\cite{wavelet_forest}.

The input text is partitioned hierarchically into
\emph{blocks}, \emph{superblocks}, and \emph{hyperblocks}. Blocks have
size $b$, superblocks have size $b_s$, and hyperblocks have size
$b_h$, where $b$ divides $b_s$ and $b_s$ divides $b_h$. Thus, each
superblock contains $b_s / b$ blocks, and each hyperblock contains
$b_h / b_s$ superblocks. In the default wavelet-forest implementation,
the superblock size is set to $2^{20}$. Technically, wavelet forests
also allow variable-sized blocks, which can improve space efficiency.
For clarity, however, we restrict the discussion here to fixed-size
blocks.

For each block, superblock, and hyperblock, we store rank information
at its left boundary, that is, at the first position of the
corresponding range. We refer to this value as the \emph{rank at the
block}, \emph{rank at the superblock}, and \emph{rank at the
hyperblock}, respectively. These values are represented by arrays
$A_b$, $A_s$, and $A_h$.
More specifically, let $c$ be a symbol. Then the rank of $c$ at
hyperblock $i_h$ is given by $A_h[c, i_h]$. If superblock $i_s$ is
contained in hyperblock $i_h$, then the rank of $c$ at superblock
$i_s$ is $A_h[c, i_h] + A_s[c, i_s]$. Similarly, if block $i_b$ is
contained in superblock $i_s$ and hyperblock $i_h$, then the rank of
$c$ at block $i_b$ is $A_h[c, i_h] + A_s[c, i_s] + A_b[c, i_b]$. In
other words, these arrays store rank values in a hierarchical,
relative form.

Each block is represented by its own wavelet tree, which can be
located via superblock and block headers. These wavelet trees are
pointerless and Huffman-shaped, and encode their blocks
independently of the rest of the text.
Moreover, because each wavelet tree is built only for the local
alphabet of its block, its height can be smaller than $\log \sigma$.

To further improve query time, the wavelet forest may also use
\emph{navigational block headers}. When enabled, each block stores
prefix-rank information for the nodes at each level of its wavelet
tree. This reduces the number of bitvector rank queries required during
traversal and can speed up traversals used by both rank and select
queries.

%% file: implementation.tex
\section{Answering Select Queries}\label{sec:implementation}

In this section, we describe how we implemented select queries on
wavelet forests. We first provide an outline of the procedure, and
then describe each step in more detail. To compute $\Select{S}{j}{c}$,
i.e., the position of the $j$th occurrence of $c$ in $S \in \Sigma^n$,
we proceed as follows:

\begin{enumerate}

\item Identify $i_h$, the index of the hyperblock containing the $j$th
  occurrence of $c$ in $S$.

\item Identify $i_s$, the index of the superblock containing the $j$th
  occurrence of $c$ in $S$.

\item Identify $i_b$, the index of the block containing the $j$th occurrence
  of $c$ in $S$. Let $j'$ denote the localized select query argument
  relative to the block. In other words, let $j' = j -
  \Rank{S}{(i_b - 1)b}{c}$.

\item Navigate the wavelet tree of that block to reach the leaf that
  corresponds to $c$ while maintaining a stack of nodes on the path to
  that leaf.

\item Navigate back up the same wavelet tree using the standard wavelet-tree
  select algorithm with $j'$ as the query argument. Let $k$ denote
  the result of this localized select query.

\item Return $(i_b - 1)b + k$.
\end{enumerate}

In the first step, we determine the hyperblock index $i_h$ by binary
searching the array of hyperblock ranks. More precisely, $i_h$ is the
largest hyperblock index such that $A_h[c, i_h] < j$, where $A_h[c, i]$
stores the rank of $c$ at the beginning of the $i$th hyperblock.

In the second step, we determine the superblock index $i_s$ by binary
searching only among the superblocks contained in hyperblock $i_h$.
Thus, $i_s$ is the largest index in the range $((i_h - 1)b_h / b_s \dd
i_h b_h / b_s]$ such that $A_h[c, i_h] + A_s[c, i_s] < j$.

In the third step, we determine the block index $i_b$ by scanning the
block headers of superblock $i_s$ from right to left. Here, a linear
scan is necessary because not every block stores the rank of every
symbol. A right-to-left scan is preferable to a left-to-right scan
because it stops as soon as it reaches the first block whose rank of
$c$ is smaller than $j$. Accordingly, $i_b$ is the largest block index
in the range $((i_s - 1)b_s / b \dd i_s b_s / b]$ such that $A_h[c, i_h]
+ A_s[c, i_s] + A_b[c, i_b] < j$. Equivalently, $i_b$ is the first
block encountered by the right-to-left scan whose rank of $c$ is
smaller than $j$. Once $i_b$ is known, we compute the localized query
value $j' = j - \Rank{S}{(i_b - 1)b}{c}$.

In the fourth step, we access the wavelet tree of block $i_b$ and
descend to the leaf corresponding to $c$. This step is needed because
the wavelet forest is pointerless, so before starting the localized
select query we must explicitly walk from the root to the correct
leaf. During this traversal, we compute the relevant node information
on the fly. The traversal is similar to the one used for rank queries,
except that it does not perform a bitvector-rank operation at each
node. Because the next step backtracks from the leaf to the root, we
store the necessary node information in a stack as we descend. The
stack can be stored in a small array. Recall that the individual wavelet
trees are Huffman-shaped. The minimum total
weight of a Huffman tree of height $h$ is attained by leaf weights $1,
F_1, F_2, \dots, F_h$, where $F_i$ is the $i$th Fibonacci number and
$F_1 = F_2 = 1$. Hence, the minimum block length that can induce
height $h$ is $1 + \sum_{i=1}^{h} F_i = F_{h+2}$. Therefore, for
blocks of size at most $2^{16}$, the height is at most $22$, because
$F_{24} = 46368 \leq 2^{16} < 75025 = F_{25}$.

In the fifth step, we perform a localized select query with argument
$j'$ on the wavelet tree of block $i_b$. This is exactly the standard
wavelet-tree select algorithm~\cite{wavelet_tree}, applied to the
canonical Huffman-shaped wavelet tree of a single block. The algorithm
uses the stack built in the previous step to recover the required node
information while backtracking from the leaf to the root. We denote
the resulting local position by $k$.

Finally, in the sixth step, we return $(i_b - 1)b + k$. Here, $(i_b -
1)b$ is the number of positions preceding block $i_b$, and $k$ is the
position of the desired occurrence within that block.

%% file: experiments.tex
\section{Experimental Results}\label{sec:experiments}

\subparagraph{Setup}

Our experiments were conducted using an Intel
Core i7-1355U CPU with 12\,MiB L3 cache, 32\,GiB of RAM,
and a 64-bit Linux Ubuntu 22.04.5 (kernel 6.8.0-60). The
frequency scaling for the CPU was set to
\texttt{performance}. Our code was compiled with \texttt{g++}
11.4.0 using the flags \texttt{-DNDEBUG -msse4.2 -funroll-loops -O3}. We measured time using
\texttt{std::chrono::high\_resolution\_clock} and the size of
data structures via serialization.

\subparagraph{Implementations}

For a fair comparison between wavelet forests and other wavelet-tree
variants, we used the implementations provided by the \texttt{SDSL}
library~\cite{sdsl}.\footnote{In preliminary experiments, we also
  examined wavelet matrices~\cite{wavelet_matrix}. However, wavelet
  matrices are designed to perform well on large alphabets, and we
  observed that, for our datasets, which have relatively small
  alphabets, wavelet trees and wavelet forests consistently provided
  better time--space trade-offs.} To clearly distinguish the different
implementations in our experiments, we use the following naming
convention:

\begin{itemize}
\item \emph{WF:} the wavelet forest~\cite{wavelet_forest}, extended
  with our implementation of select queries.\footnote{The default
  superblock size is $2^{20}$.}
\item \emph{WT-HF:} the Huffman-shaped wavelet
  tree~\cite{wavelet_tree} provided by SDSL~\cite{sdsl}.
\item \emph{WT-RL:} the run-length-compressed wavelet tree available
  in SDSL~\cite{sdsl,rlwt}.
\end{itemize}

Both WF and WT-HF store their internal data in bitvectors, and each of
them can be instantiated with any of the following bitvector
representations:

\begin{itemize}
\item \emph{BV:} the uncompressed bitvector implementation in SDSL.
\item \emph{HYB:} the hybrid bitvector~\cite{hybrid_bitvector}, using
  superblock rates 8, 16, 32, and 64 (default: 16).\footnote{We use the hybrid
  bitvector augmented with support for select
  queries~\cite{hybridselect}.}
\item \emph{RRR:} the RRR bitvector~\cite{rrr}, with
  block sizes 15, 31, 63, 127 (default: 63).
\end{itemize}

We refer to a particular combination of wavelet-tree variant and
bitvector type by concatenating their aliases with a dash. For
example, WF-HYB denotes the wavelet forest instantiated with hybrid
bitvectors.

\subparagraph{Datasets}

Our dataset consists of two parts. The first part contains four
non-repetitive and four repetitive texts from the Pizza\&Chili
corpus~\cite{pizzachili}.
The second part consists of four 8\,GiB
texts from various sources: the 1000 Genomes
Project~\cite{1000genomes}, Common Crawl~\cite{commoncrawl},
Wikipedia, and Project Gutenberg~\cite{gutenberg}.
Statistics for all texts are given in \cref{tab:dataset}.

\begin{table*}[t!]
  \centering
  \begin{minipage}[c]{0.49\textwidth}
    \centering
    \begin{tabular}{lrrr}
      \hline
      Name            & $\sigma$ & $n/2^{20}$ & $n/r$ \\
      \hline
      dna             & 16       & 200        & 1.63  \\
      english         & 225      & 200        & 2.91  \\
      sources         & 230      & 200        & 4.40  \\
      dblp.xml        & 96       & 200        & 7.09  \\
      \hline
      para            & 5        & 410        & 27    \\
      world\_leaders  & 89       & 44         & 82    \\
      kernel          & 160      & 246        & 93    \\
      einstein.en.txt & 139      & 446        & 1611  \\
      \hline
    \end{tabular}
  \end{minipage}
  \hfill
  \begin{minipage}[c]{0.49\textwidth}
    \centering
    \begin{tabular}{lrrr}
      \hline
      Name        & $\sigma$ & $n/2^{20}$ & $n/r$ \\
      \hline
      1000genomes & 51       & 8000       & 2.29  \\
      commoncrawl & 234      & 8000       & 5.73  \\
      enwiki      & 212      & 8000       & 3.82  \\
      gutenberg   & 216      & 8000       & 2.69  \\
      \hline
    \end{tabular}
  \end{minipage}

  \vspace{5mm}

  \caption{Statistics of the datasets used in our experiments,
    including alphabet size $\sigma$, size in MiB ($n / 2^{20}$), and
    average run length in the BWT ($n/r$).}\label{tab:dataset}
\end{table*}

\subsection{Experiment 1: Performance of Select Queries}\label{sec:experiments-exp-1}

In the first experiment, we compare the performance of wavelet forests
for select queries against other wavelet-tree variants. We also vary
the bitvector representation and its parameters. For each text in our
dataset, we compute its BWT~\cite{bwt} and generate $10^{5}$ random
select queries. For each variant, we measure average
query time and total index size.

\begin{figure*}[p]
  \centering
  \includegraphics[page=1,width=\textwidth]{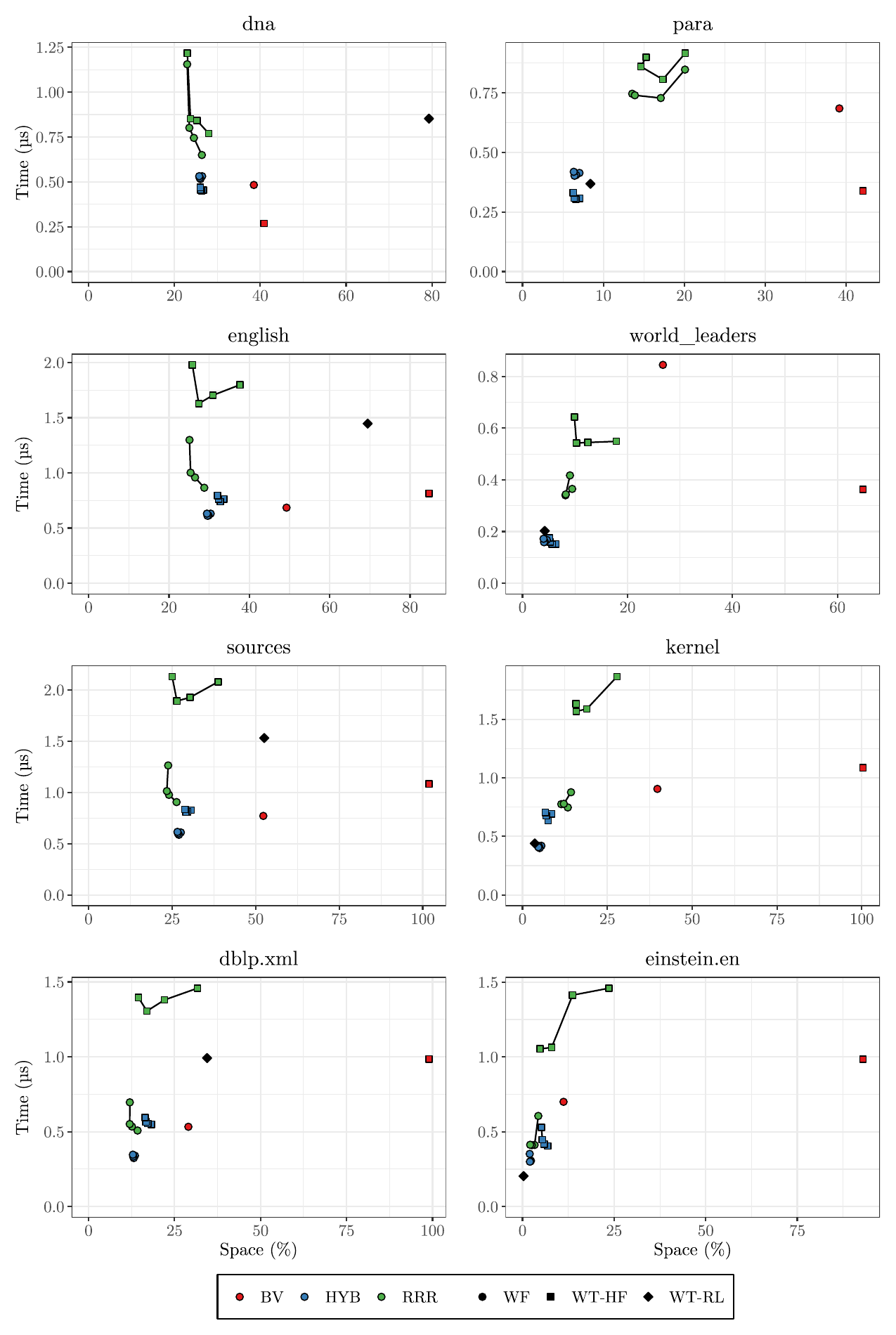}
  \caption{Time-space trade-offs for select queries on texts from the
    Pizza\&Chili corpus. Time is measured in microseconds, and space
    is expressed as a percentage of the original text size.
    Non-repetitive and repetitive texts are shown in the left and
    right columns, respectively.}\label{fig:select}
\end{figure*}

\begin{figure*}[t]
  \centering
  \includegraphics[page=1,width=\textwidth]{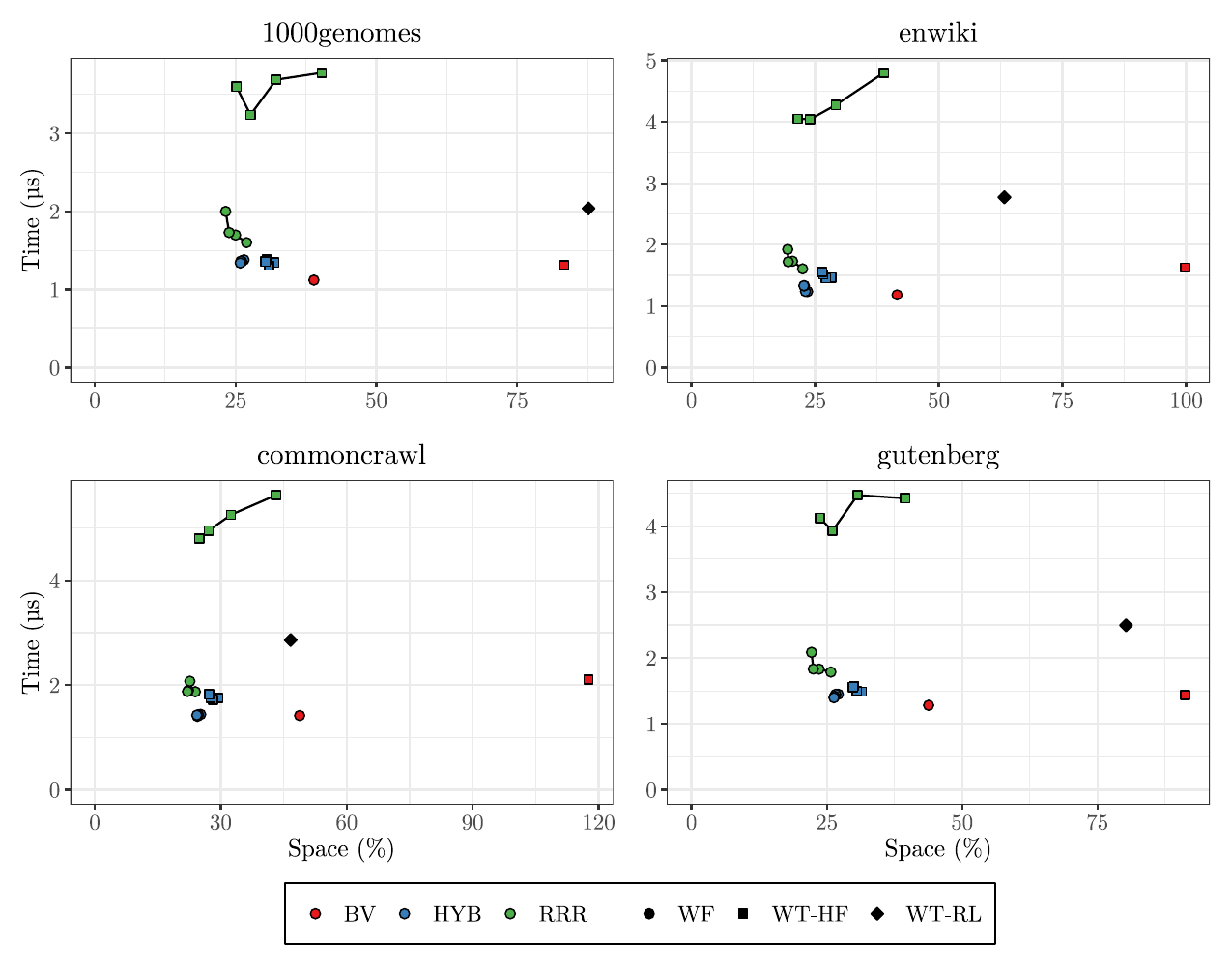}
  \caption{Time-space trade-offs for select queries on 8\,GiB texts.
    Time is measured in microseconds, and space is expressed as a
    percentage of the original text size.}\label{fig:select-8G}
\end{figure*}

The results for the Pizza\&Chili texts are shown in
\cref{fig:select}. In the majority of cases, wavelet forests are both
faster and more space-efficient than standalone wavelet trees. When
combined with RRR bitvectors, wavelet forests improve query time by up
to a factor of two while using the same or less space. Notable
exceptions occur for the \texttt{para} and \texttt{world\_leaders}
texts, whose small alphabets limit the ability of wavelet forests to
exploit the difference between local and global alphabet sizes. A
particularly effective combination is obtained by pairing wavelet
forests with hybrid bitvectors. This variant achieves strong overall
performance on both non-repetitive and highly repetitive texts, making
it a good off-the-shelf choice in practice. The results for the larger
texts, shown in \cref{fig:select-8G}, exhibit similar trends.

\subsection{Experiment 2: Effect of Superblock Size}\label{sec:experiments-exp-2}

In the second experiment, we study how the superblock size affects the
space usage and select-query time of the wavelet forest. We used two
representative texts from the Pizza\&Chili corpus: \texttt{english},
representing non-repetitive data, and \texttt{kernel}, representing
repetitive data. As in the previous experiment, we first compute the
BWT of each text and then run random select queries. For HYB and RRR,
we used the default parameter settings.

\begin{figure*}[t]
	\centering
	\includegraphics[page=1,width=\textwidth]{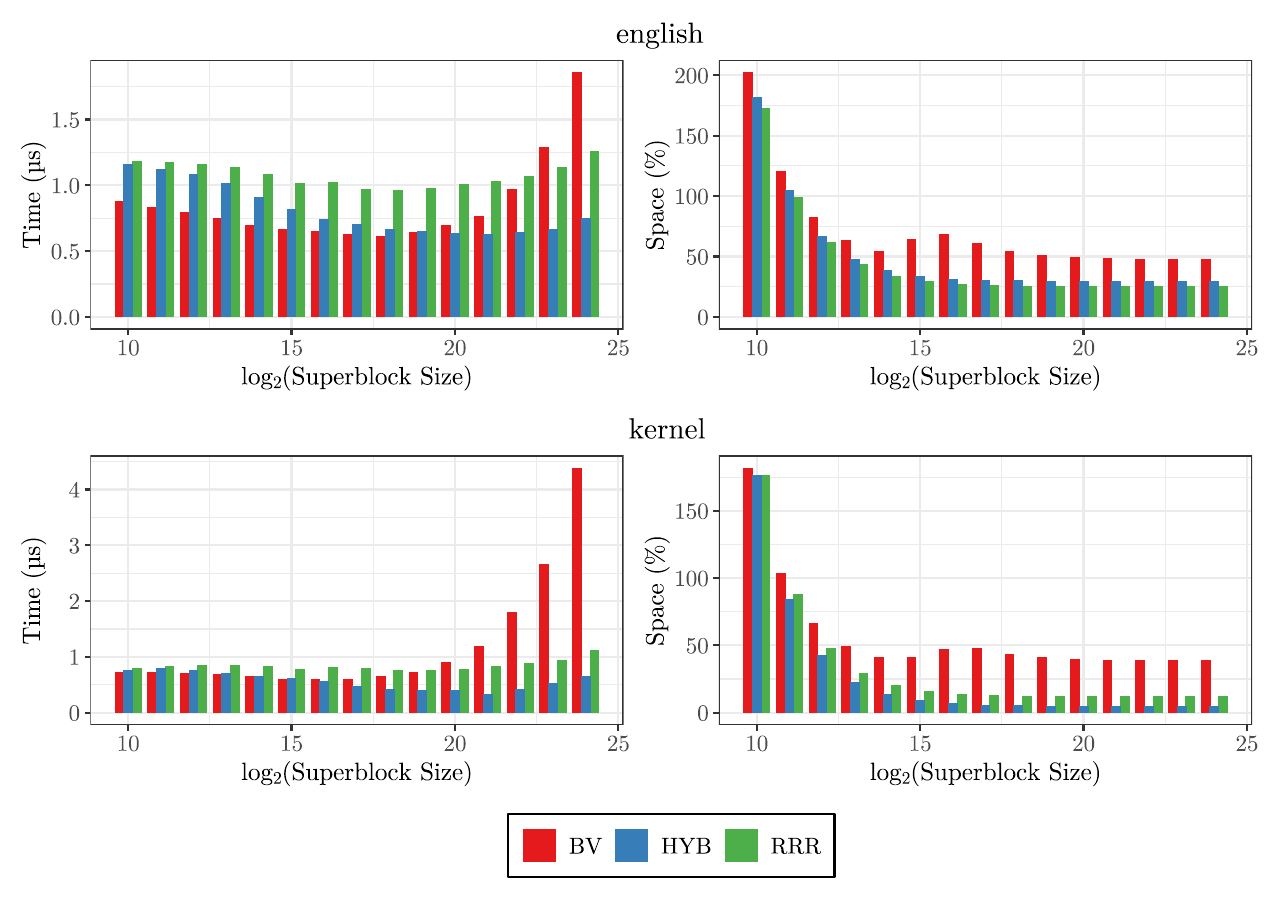}
	\caption{Effect of superblock size on the select-query performance
		of wavelet forests. Time is measured in microseconds, and space
		is expressed as a percentage of the original text size.}
	\label{fig:superblock-size}
\end{figure*}

The results are shown in \cref{fig:superblock-size}. They indicate
that the superblock size can have a substantial effect on performance.
If the superblock size is too large, query time can deteriorate.
On the other hand, if the
superblock size is too small, space usage increases significantly.
Overall, our experiments confirm that choosing
a superblock size around $2^{20}$ is a robust practical default.
The same value was previously observed to provide a good trade-off
for rank queries in wavelet forests~\cite{wavelet_forest}.

\subsection{Experiment 3: Effect of Navigational Block Headers}\label{sec:experiments-exp-3}

In the third experiment, we evaluate how navigational block headers
affect the space usage and select-query performance of wavelet
forests. Navigational headers are optional components of the block
headers in the wavelet-forest implementation that speed up traversal
within the wavelet tree of the current block; see
\cite{wavelet_forest}. They are enabled by default, and the goal of
this experiment is to determine how much they help for select queries.
As in the previous experiment, we used the BWTs of two representative
texts from the Pizza\&Chili corpus, namely \texttt{english} and
\texttt{kernel}.

The results are shown in \cref{tab:experiment-3}. For
\texttt{english}, across all bitvector types, navigational block
headers improve query time by about $1.1\times$ to $1.72\times$,
while increasing space by at most $1.1\%$. For \texttt{kernel},
the corresponding speedups range from about $1.02\times$ to
$1.52\times$, with space overhead up to $2.4\%$. These results
show that navigational block headers substantially improve query time
at the cost of only a very small increase in space.

\begin{table*}[t]
	\centering
	\caption{Effect of navigational block headers on wavelet-forest
		select-query performance. Time is measured in microseconds, and
		space is expressed as a percentage of the original text size.}
	\begin{tabular}{llrrrr}
		\hline
		\multirow{2}{*}{Text}    & \multirow{2}{*}{Bitvector} & \multicolumn{2}{c}{Space (\%)} & \multicolumn{2}{c}{Time ($\mu$s)}                    \\
		                         &                            & With                           & Without                           & With   & Without \\
		\hline
		\multirow{3}{*}{english} & BV                         & 49.2547                        & 48.7405                           & 0.6913 & 0.7658  \\
		                         & HYB                        & 29.8543                        & 29.7226                           & 0.6215 & 0.7518  \\
		                         & RRR                        & 25.3844                        & 25.2495                           & 0.9965 & 1.7224  \\
		\hline
		\multirow{3}{*}{kernel}  & BV                         & 39.7453                        & 38.8623                           & 0.9047 & 0.9223  \\
		                         & HYB                        & 5.0468                         & 4.9298                            & 0.4018 & 0.5021  \\
		                         & RRR                        & 12.1714                        & 12.0439                           & 0.7804 & 1.1842  \\
		\hline
	\end{tabular}
	\label{tab:experiment-3}
\end{table*}